# Su-Schrieffer-Heeger Model Inspired Acoustic Interface States and Edge States

Xin Li[1,a], Yan Meng[1,a], Xiaoxiao Wu[2,a], Sheng Yan[2], Yingzhou Huang[1], Shuxia Wang[1,b], Weijia Wen[2,b]

[1]*Chongqing Key Laboratory of Soft Condensed Matter Physics and Smart Materials, College of Physics, Chongqing University, Chongqing, 400044, China*

[2]*Department of Physics, The Hong Kong University of Science and Technology, Clear Water Bay, Kowloon, Hong Kong, China*

[a] These authors contributed equally to this work.

[b] Correspondence and requests for materials should be addressed to S.X.W. (email: wangshuxia@cqu.edu.cn) or to W.J.W. (email: phwen@ust.hk).

**ABSTRACT**

If a full band gap closes and then reopens when we continuously deform a periodic system while keeping its symmetry, a topological phase transition usually occurs. A common model demonstrating such a topological phase transition in condensed matter physics is the Su-Schrieffer-Heeger (SSH) model. As well known, two distinct topological phases emerge when the intracell hopping is tuned from smaller to larger with respect to the intercell hopping in the model. The former case is topologically trivial, while the latter case is topologically non-trivial. Here, we design a 1D periodic acoustic system in exact analogy to the SSH model. The unit cell of the acoustic system is composed of two resonators and two junction tubes connecting them. We show that the topological phase transition happens in our acoustic analog when we tune the radii of the junction tubes which control the intercell and intracell hoppings. The topological phase transition is characterized by the abrupt change of the geometric Zak phase. The topological interface states between non-trivial and trivial phases of our acoustic analog are experimentally measured, and the results agree very well with the numerical values. Further, we show that topologically non-trivial phases of our acoustic analog of the SSH model can support edge states, on which the discussion is absent in previous works about topological acoustics. The edge states are robust against localized defects and perturbations.

**INTRODUCTION**

The field of topological physics is growing rapidly in condensed matter physics, from quantum Hall effect[1] to topological insulators[2] and Weyl semimetals[3]. Topological insulators possess topologically non-trivial band gaps and support one-way surface states. Berry phase[4-7] is usually utilized to characterize the topological phase in both classical and quantum system[8-12]. In three-dimensional (3D) system, two bands linearly intersect each other at the Weyl point[13], and a Weyl point is a source or drain of Berry flux, that is, a topological charge characterized by the surface integral of its Berry curvature[14,15]. In two-dimensional (2D) system, the surface integral of Berry curvature is indicated by a Chern number[16]. In addition, for one-dimensional (1D) system, the integral of Berry connection is defined as a Zak phase [17].

In the 1D system, one of the most representative models with topologically non-trivial phases is the Su-Schrieffer-Heeger (SSH) model[18]. This model describes the electrons' staggered hopping in the joint of unit cells, and there are two distinguished kinds of hopping defined as intracell hopping and intercell hopping in this model[18-25]. By controlling the hopping amplitude, energy band gap of the 1D chains can close and reopen which implies the topological phase transition[26]. For instance, in the 1D system, if the strength of intercell hopping is stronger than that of intracell hopping, this system possesses a topological non-trivial phase. Otherwise, if the strength of intracell hopping is stronger than that of intercell hopping, the system possesses a topological trivial phase. When the 1D lattice with topologically non-trivial phase, topological edge states can be observed[25,27]. Moreover, topological interface states emerge when two lattices with different topological phases are connected [18]. Inspired by the SSH model, the topological phase in the 1D system has been extended to various classical wave systems, such as mechanical lattice[28,29], photonic crystal[30] and phononic crystal[31].

Phononic crystals are artificial materials that can periodically modulate acoustic waves and possess acoustic band gaps. The band structures of phononic crystals can be tuned by changing the geometric parameters. Specifically, the acoustic band gaps can be tuned from closed to reopen, and the band inversion implies topological phase transitions[17,30,32,33]. Topological acoustics in higher dimensional systems[34-40] and 1D systems[30,41] have been frequently discussed. Phononic crystals in the audible-sound regime usually have macroscopic dimensions so they can be adjusted easily. As a consequence, the analogs of topological insulators in the acoustic system can be designed and fabricated very conveniently and may have potential applications in monochromatic sound generator [42].

In this work, we experimentally validate a kind of 1D phononic crystal whose topological properties can be mapped to the SSH model. The phononic crystal is composed of cylindrical waveguides and resonators with periodically alternating structures (proposed by Z. Yang and B. Zhang in Fig. 1 of Ref.40). Topological interface states are realized by connecting two 1D phononic crystals with different topological phases. The existence of topological interface states is predicted by topological phase transition and demonstrated by simulations and experiments. The measured results agree well with the simulated ones. In addition, topological edge states can be realized by a single topologically non-trivial phononic crystal. The existence of topological edge states are predicted by the topological phase of each band and observed in simulations.

**RESULTS AND DISCUSSION**

The experimental setup and sample are shown in Fig. 1(a). The sample consists of two kinds of phononic crystals, which are both fabricated by a 3D printer using photopolymer resin. The magenta arrow indicates the interface between the two kinds of phononic crystals. A unit cell of the phononic crystal on the left (right)-hand side of the interface is marked by red (green) dished rectangle and referred to unit-A (B).

Each unit cell consists of two identical and vertically oriented cylindrical cavities acting as the resonators and two horizontally oriented cylindrical tubes with different radii acting as the junctions. Each kind of the phononic crystal is analogous to the SSH model, and the vertical (horizontal) cylindrical cavities corresponding to the site (bond). The sketch of unit-A is illustrated in Fig. 1(b), and the dimensions are $a=200$ mm, $t=5$ mm, $r=40$ mm, $h=80$ mm, $w_1=13$ mm, and $w_2=20$ mm. The sketch of unit-B is illustrated in Fig. 1(c), and the geometric parameters are the same as unit-A, except for $w_1=20$ mm, $w_2=13$ mm. As depicted, the intercell hopping is controlled by the wave-guiding junction tube with radius $w_2$ in the center of the unit cell, while the intracell hopping is controlled by the wave-guiding junction tube with radius $w_1$ near the boundaries of the unit cell. The system is filled with air, and the thickness of the walls of the samples $t$ is large enough (> 2 mm) to be regarded as rigid boundaries.

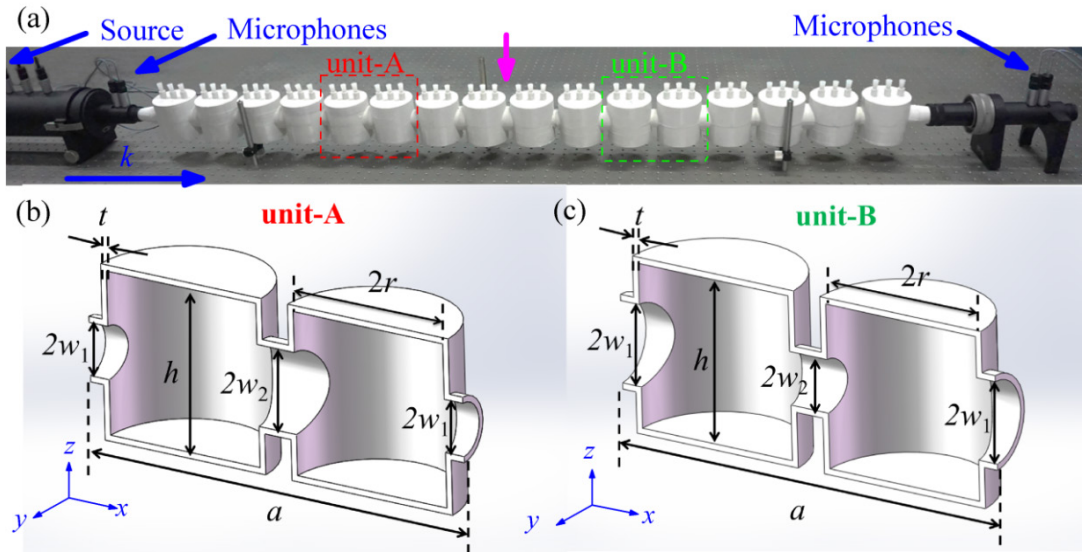

Fig. 1. (a) Experimental set-up. The experimental system consists of a loud speaker and four microphones. The sample is an acoustic waveguide which consists of two kinds of phononic crystals. These phononic crystals are connected, and the junction

is marked by a magenta arrow. The phononic crystal on each side of the junction possesses a unit cell of either unit-A or unit-B, and the unit cells on each side are marked by red and green dashed boxes, respectively. Acoustic waves are generated by the loud speaker and propagate through the acoustic system in the direction of the wave vector *k* (indicated by a blue arrow). (b)-(c) Cross sections of unit-A and unit-B, and their geometric parameters are correspondingly labeled. The dimensions for unit-A are *a* = 200 mm, *t* = 5 mm, *h* = 80 mm, *r* = 40 mm, $w_1$ = 13 mm, $w_2$ = 20 mm. The dimensions for unit-B are all the same with unit-A, except $w_1$ = 20 mm, $w_2$ = 13 mm.

The eigenmodes and band structures of the phononic crystals can be simulated by the finite element method (COMSOL Multiphysics). During simulations, the Bloch-Floquet boundary conditions are used on the periodic faces of each unit cell, and the sound hard boundary conditions are applied on the other boundaries. The band diagrams of unit-A and unit-B are shown in Fig. 2(a) and Fig. 2(c), respectively. The band structure of unit-C with the parameters of $w_1 = w_2 = 16.5$ mm are shown in Fig. 2(b), where the band crossing happens. It can be observed that the band structures of unit-A and unit-B are the same because these two unit cell corresponding to the same structure with a different choice of inversion center. The eigenmodes of the band edge states marked in Figs. 2(a) and 2(c) are correspondingly shown in Fig. 2(d) and Fig. 2(e). It can be observed from Figs. 2(d) and 2(e) that the acoustic field pressure of the eigenmodes varying along the wave propagation direction (*x*-direction), and almost uniform in the z-direction. All the eigenmodes are symmetric or anti-symmetric with respect to the inversion center. The Zak phase for each band can be determined by the symmetry of these eigenmodes. For instance, in Fig. 2(a) the symmetry of the eigenmodes for the lower and upper edges of the second band is the same, thus the Zak phase is 0 [17,26,30]. Similarly, in Fig. 2(c) the symmetries of the eigenmodes for the lower and upper edges of the second band are different, thus the Zak phase is π.

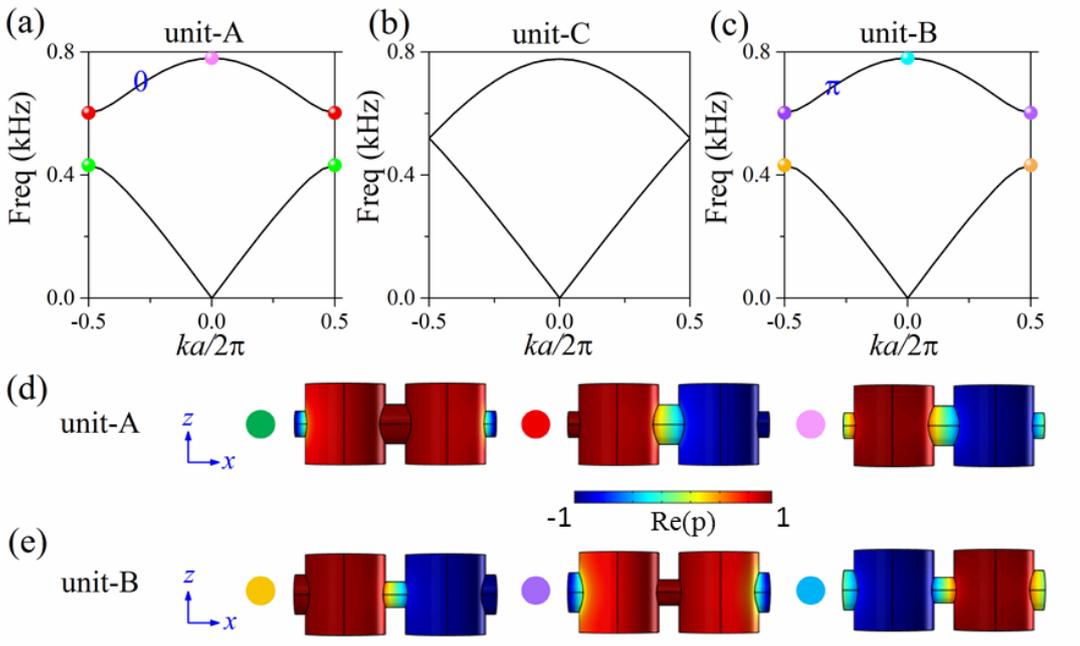

Fig. 2. Calculated band structures of the unit cells with $w_1$=13 mm, $w_2$=20 mm (a), $w_1$ = $w_2$ =16.5 mm (b), $w_1$ = 20 mm, $w_2$ = 13 mm (c). The unit cells are referred to as unit-A, unit-C, and unit-B, respectively. The Zak phases of each frequency band are marked by blue numbers, 0 or π. The eigenmodes corresponding to the colored points marked in the frequency bands of unit-A and unit-B are shown in (d) and (e), respectively, where the color scale indicates the normalized real part of acoustic pressure fields. It can be seen that the pressure field of each eigenmode is either symmetric or anti-symmetric with respect to the center plane of unit-A or unit-B. The symmetries of eigenmodes at the boundaries or the center of a certain frequency band in the reciprocal space are important to predict its Zak phase.

To confirm the values of the Zak phase determined from the field symmetry, we also numerically calculated their values. We adopt the sign convention $e^{i\omega t}$ for harmonic time dependence in this work. The Zak phase for each band can be defined as [43]:

$$\theta_{Zak} = i\int_{-\pi/a}^{\pi/a} \langle u_k | \partial_k | u_k \rangle dk, \qquad (1)$$

where $k$ is the Bloch quasi-momentum, $u_k = e^{-ikx}\psi_k(x)$ is the periodic Bloch's function, $\psi_k(x)$ is the Bloch wave, and $a$ is the period. More specifically, the Zak phase for the $n$-th isolated band of our acoustic system can be expressed as [44-47]:

$$\theta_n^{Zak} = \int_{-\pi/a}^{\pi/a} dk \left[ i\int_{\text{unit cell}} \frac{1}{2\rho v^2} dxdydz u_{n,k}^*(x,y,z)\partial_k u_{n,k}(x,y,z) \right], \qquad (2)$$

where $\theta_n^{Zak}$ represents the Zak phase of the $n$-th band, $\rho$ is the density of the air, $v$ is the speed of sound in air, and $u_{n,k}(x,y,z)$ is the periodic-in-cell part of the normalized Bloch eigenfunction of the state on the $n$-th band with wave vector $k$. The Zak phase in our system can be any value if we use an arbitrary choice of the unit cell [17,43]. However, if the chosen unit cell possesses mirror symmetry, the value of the Zak phase is equal to either 0 or $\pi$. The calculated Zak phase (0 or $\pi$) for a mirror symmetric system still depends on the choice of the origin, and in this work, the origin is at the center of the unit cells depicted in Figs. 1 and 2. The calculated results confirm the previous predictions based on the symmetry of the band edge states.

The topological properties of the phononic crystal in this paper can be modulated by the geometric parameters of each unit cell. Here, we change the difference of the radius for two horizontal junction tubes δw (δw = $w_1 - w_2$) to realize the topological phase transition. In order to observe the evolution of eigenmodes in the parameters space, the eigenfrequnency of each band edge state is illustrated in Fig. 3. As shown in Fig. 3, the black line and the red line denote symmetric and anti-symmetric band-edge modes, respectively. It is obvious that frequency band closes and reopens as δw increases, and the critical point δw = 0 is the band inversion point. The topological phase of each band gap can be determined by the summation of Zak phases of all bulk bands below the band gap [42,48]. The topological phases of the 1st band gap are marked in cyan (Zak phase 0) and in magenta (Zak phase π). The symmetric distribution of the upper and lower band edges is due to the chiral symmetry in our system near the boundaries of the Brillouin zone (see Supplementary Note 1) [49].

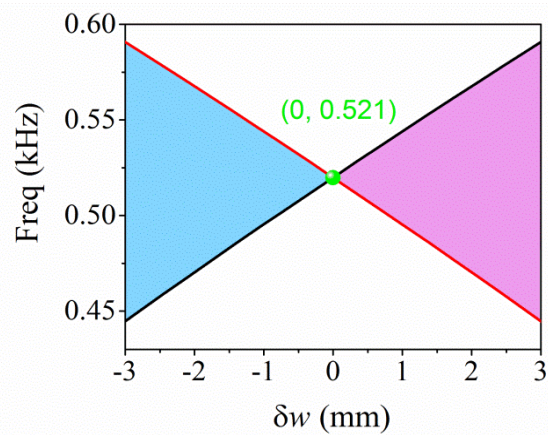

Fig. 3. Band edges of the first two eigenmodes as a function of δw = $w_1 - w_2$, while the average of $w_1$ and $w_2$ is kept as 16.5 mm. The red and black lines correspond to anti-symmetric and symmetric band-edge states, respectively. As δw increases from negative to positive, the first band gap closes and reopens, indicating a band inversion process. The band gaps with different topological properties are filled with different colors. The green point indicates the crossing position (0, 0.521) of the eigenmodes. Thus, any two phononic crystals with δw on different sides of the green point (δw = 0) possess different topological properties in the first band gap.

Topological interfaces can be achieved by connecting two phononic crystals with different topological phases in certain band gaps [50,51]. The two segments of phononic crystals are connected with a junction tube of radius 20 mm (see Supplementary Note 2). Here, we chose a set of suitable values of δw to constitute phononic crystals with different topological phases and connect them to realize topological interface states in the first band gap. In Fig. 1(a), the radius difference for the left-side phononic crystal is chosen to be δw = –7 mm and for the right-side one is chosen to be δw = 7 mm (this whole structure is referred as $S_1$). The eigenmodes of $S_1$ are shown in Fig. 4(d). For comparison, another set values of δw are chosen to be δw = –7 mm and δw = –3 mm on the left and right hand, respectively (this whole structure is referred to as $S_2$). The eigenmodes of $S_2$ are shown in Fig. 4(a), it can be observed that there exist eigenmodes with eigenfrequency in the middle of the band gap (indicated by red line). This comparison proves that topological interface states emerging when the topological phase transition occurs, which implies the topological phases for the two phononic crystals on either side of the junction are different.

In addition, the transmission spectra of $S_2$ are simulated and measured, and the results are shown in Fig. 4(b) and Fig. 4(c), respectively. The corresponding results for $S_1$ are shown in Fig. 4(e) and Fig. 4(f). Comparing the transmission spectra of $S_1$ and $S_2$, an obvious transmission peak can be observed in either the simulated or the measured transmission spectra for $S_1$. As shown in Fig.4, the experimental transmission spectra reproduce the simulated one quite well, and the deviation is less than 3%. The lower amplitude for the measured transmission spectra is mainly caused by friction dissipations and also leakages of sound from the sample, which is hard to avoid since the plastic boundaries are not strictly rigid (see Supplementary Note 4). In summary, the existence of topological interface states has been soundly proven and observed by the eigenfrequency spectra and the transmission spectra.

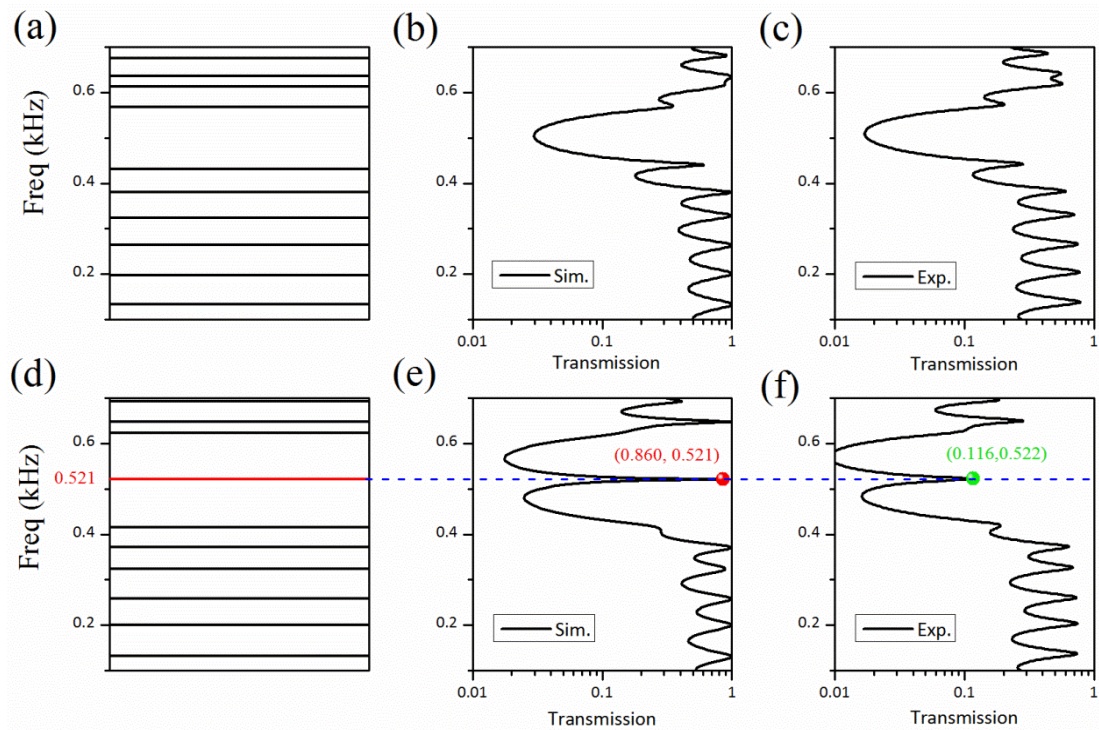

Fig. 4. (a) The eigenfrequency spectra of the acoustic waveguide which consists of four units-A ($\delta w = -7$ mm) and four units-A' ($\delta w = -3$ mm) on each side of the junction. An obvious band gap appears in the vicinity of 0.5 kHz of the eigenfrequency spectra, corresponding to the first band gap of this acoustic waveguide. As the topological phase for the structures on each side of the junction is the same, so there is no topological interface state in the first band gap. Simulated (b) and measured (c) transmission spectra for the structure described in (a), an obvious gap can be observed in the vicinity of 0.5 kHz in these two transmission spectra. (d) Eigenfrequency spectra of the acoustic waveguide consists of four units-A ($w = -7$ mm) and four units-B ($w = 7$ mm). A frequency level (red) appears in the first band gap, which demonstrates the existence of topological interface states in this system. (e) Simulated transmission spectrum for the structure described in (b). A sharp transmission peak arises in the first band gap, and the frequency of the transmission peak (0.521 kHz) is marked by a red dot. (f) Measured transmission spectrum for the structure described in (d). A sharp transmission peak arises in the first band gap, and the frequency of the transmission peak (0.522 kHz) is marked by a green dot.

To further verify the existence of the interface state in the first band gap of $S_1$, the special field distributions of acoustic pressure for the interface state is measured. The simulated field distributions of acoustic pressure for the interface state are shown in Fig. 5(a). We make many periodic holes on the samples to probe the internal acoustic field by inserting the microphones into the holes. When measuring the acoustic field at some of the holes, all the other holes are sealed with rubber plugs to prevent sound leakage. The measured spatial distribution of acoustic pressure field is shown by red circles in Fig. 5(b) and the simulated field distribution along the central line of the sample (indicated by the red dashed line in Fig. 5(a)) is shown as the black solid line for comparison. It can be observed in Fig. 5(b) the profile of the acoustic pressure for the interface state decays exponentially away from the interface junction ($x=0$). Excellent agreement has been achieved between the measurement and the simulation, and it confirms that the acoustic pressure is localized in the vicinity of the interface which is a key indicator of interface states.

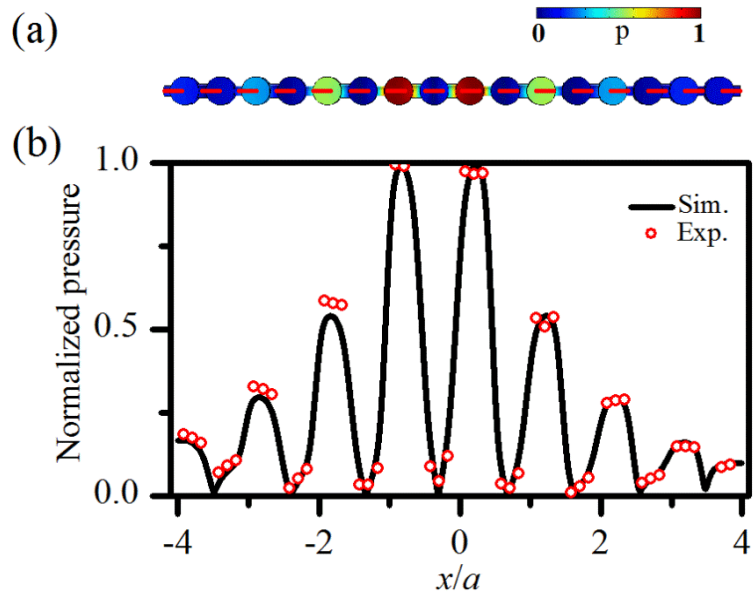

Fig. 5. (a) The simulated spatial distribution of the acoustic pressure field for the interface state in $S_1$. The red dashed line indicates the center line of the structure. The color scale represents the normalized amplitude of sound pressure. (b) The measured acoustic pressure field for the interface state of $S_1$ is indicated by red circles, and the simulated field distribution along the red dashed line is indicated by black solid line. The amplitude is normalized by the maximum value of $P_0$.

Analogous to the SSH model, except for the topological interface states realized by connecting two phononic crystals with different topological phases, the topological edge states at the ends of phononic crystals of topologically non-trivial phases can be obtained as well [25]. To confirm the existence of topological edge states, we numerically simulated the eigenfrequency spectra of a phononic crystal composed of 29 unit cells. The simulated eigenfrequency spectra of phononic crystals with 29 units-A ($\delta w = -7$ mm, referred to as PC1 below) and 29 units-B ($\delta w = 7$ mm, referred to as PC2 below) are shown in Figs. 6(c) and 6(d), respectively. The corresponding geometric illustrations are shown in Figs. 6(a) and 6(b), respectively. Each of the phononic crystals is terminated with an additional resonator on either end (blue), which makes the phononic crystals contains 60 resonators in total. As shown in Fig. 6(c), there is no edge state appears in the first band gap of PC1, as this structure possesses a topological trivial geometric Zak phase (0). On the contrary, topological edge states (red line) appear in the first band gap of PC2 at 0.516 kHz in Fig. 6(d), as this structure possesses a topological non-trivial geometric Zak phase ($\pi$). The simulated acoustic pressure fields of two degenerate topological edge states are shown in the upper and lower panel of Fig. 6(e). As shown in Fig. 6(e) the simulated pressure fields of these two topological edge states are localized at either end of PC2.

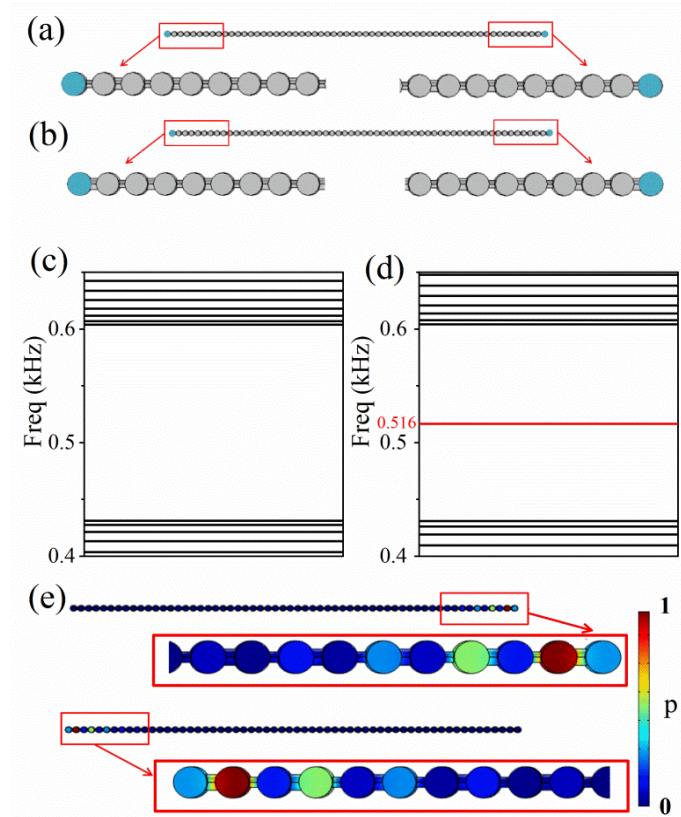

Fig. 6. (a) The phononic crystal is composed of 29 units-A, referred to as PC1. (b) The phononic crystal is composed of 29 units-B, referred to as PC2. Notice that the resonators marked as blue are additional resonators to provide a similar acoustic environment for possible edge states. (c) The simulated eigenfrequency spectrum of PC1 illustrated in (a). (d) The simulated eigenfrequency spectrum of PC2 illustrated in (b). By comparison, two degenerate edge states marked by a red line arise in (d) at 0.516 kHz. (e) Distributions of the acoustic pressure field corresponding to the red line in (d). Both of their acoustic pressure fields are localized in the vicinity around the edges. The color scale represents the normalized amplitude of acoustic pressure.

It should be noticed that the function of the terminal resonators is to provide a similar acoustic environment for the bulk of the phononic crystals. It can be proved that the existence of the topological edge states is topologically protected and does not depend on the geometry of the terminal resonators, only if the edge states are "pushed" into the bulk bands. To validate this point, the geometric structures of the terminal resonators are perturbed, and the corresponding calculated eigenfrequency spectra are shown in Supporting Information (see Supplementary Note 5). It can be seen, no matter how the terminal resonators are perturbed, no edge state appears in the first band gap of PC1. On the contrary, the existences of topological edge states in the first band gap are robust against perturbations for PC2. Therefore, the fully dimerized limits of the SSH model can be achieved in our system, and the edge states can be realized in our acoustic system. We have also calculated the localized density of states (LDOS) of the interface states and edge states, and the results are shown in supporting informing (see Supplementary Note 6)

**CONCLUSIONS**

In summary, we propose and demonstrate an analog of the SSH model in the acoustic system. The existence of topological interface states can be predicted by the topological phase of each phononic crystals, and the existence of topological interface states are verified in experiments and simulations. Meanwhile, the existence of topology edge state pairs is verified in simulations. The localized acoustic pressure field with a certain eigenfrequency in the first band gap which caused by topological interface states or topological edge states may have potential applications in the monochromatic sound generator. Meanwhile, as a result of the localized sound field in the interface or the edge, the structure may be used in acoustic devices, such as sonars, acting as a strong source [52].

## SUPPLEMENTARY MATERIAL

See Supporting Information for chiral symmetry of the acoustic analog, the details around the interface, robustness of the interface states against local defects, effect of thermal losses and leakages on transmissions, the impact of terminal resonators on edge states, and LDOS of interface states and edge states.

## ACKNOWLEDGMENT

This work was supported by an Areas of Excellence Scheme grant (AOE/P-02/12) from Research Grants Council (RGC) of Hong Kong, Fundamental Research Funds for the Central Universities (2018CDXYWU0025, 2018CDJDWL0011), Key Technology Innovation Project in Key Industry of Chongqing (cstc2017zdcy-zdyf0338), the Science and Technology Research Program of Chongqing Municipal Education Commission (KJQN201800101), and sharing fund of Chongqing University's large-scale equipment. Dr. Sheng Yan is the recipient of the 2018 Endeavour Research Fellowship funded by the Australian Department of Education and Training.

Physical Review Letters **120** (11), 114301 (2018).